# Sn whisker growth mitigation by using NiO sublayers

Vamsi Borra, Srikanth Itapu, and Daniel G. Georgiev

Department of Electrical Engineering and Computer Science, University of Toledo,

Toledo, OH 43606, U.S.A.

**Email: Daniel.Georgiev@utoledo.edu**

## ABSTRACT

The potential of NiO sublayers for whisker growth mitigation has been examined. A thin NiO film was applied on a Cu-coated substrate before the deposition of a thicker Sn layer. The growth of Sn whiskers was then followed by optical and scanning electron microscopy and was compared with the whisker growth on a control sample without the NiO sublayer. No whiskers were observed on the sample with the NiO layer even after 12 months, whereas the control sample developed whiskers of size and density that is generally expected, based on the vast amount of published work on the topic. The mechanisms of whisker growth and whisker growth suppression are briefly discussed as well.

## 1. Introduction

Sn-based solders that are extensively used in electronic manufacturing often develop electrically conductive hair-like structures on their surfaces, which are referred to as metal whiskers (MWs). These whiskers form in a rather unpredictable fashion and can have lengths ranging from a few micrometers to several millimeters. MWs can lead to current leakage and short circuits in electronic equipment causing significant losses in the automotive, airspace and other industries[1,2]. Other metals and metal alloys, such as Zn, In, Al, Au, Ag, Cd, and Pb[3,4] are also known to form whiskers. Although whiskers were discovered almost 70 years ago[5], the basic mechanism of formation is not well understood. The dominant view today appears to be that MWs' formation is due to stress relieving phenomena[6–8] but there are other hypotheses[4,9] as well, including a recent one referred to as electrostatic theory[10]. The latter provides some quantitative estimates of MW nucleation and growth rates, and their statistical distributions, which are found to be consistent with observations[11,12]. It proposes that imperfections on metal surfaces can result in small patches of net positive or negative electric charge, leading to the formation of local electric field($E$) which is mostly normal to the surface and can cause the growth of whiskers.

Various techniques for fabrication of Sn and other metal coatings[13], combined with post-deposition treatment, have been studied as methods for whisker growth mitigation[14]. Adding a small fraction of Pb (2-3%) is known to significantly suppress whiskers formation[4]. Even though this approach does not entirely eliminate the problem, it works well and has been widely used in the industry over the last several decades. Unfortunately, recently introduced laws and standards (such as RoHS and others[15]) require the elimination or the significant reduction of the use of Pb, and this *has exacerbated the problem of whiskers growth*, while a consistent and universal solution is still elusive[4].

As a more specific example, related to this work, using a Ni sublayer between a Cu substrate and a Sn layer to mitigate whiskers has shown potential in whiskers growth mitigation[16]. However, other research[17] data show that the sublayer was able to prevent whisker growth only for 6 months. Another group[18] reported the growth of whiskers even with Ni underlying layer within a relatively short duration of 2000 hrs. The reliability of the Ni sublayer approach was further questioned by the fact that whiskers were found on multilayer ceramic chip capacitors and connectors by several other research groups[19–21]. Although the Ni sublayer does not guarantee the elimination or even any substantial suppression of whiskers growth, many electronics manufacturers still use Ni as a sublayer[4].

This work is motivated in part by the fact that the Ni sublayers have indeed shown some effect on whiskers growth when compared with the Sn-coatings without a sublayer[21]. Theoretical work[4,14,22] suggests that the Ni sublayer in the Cu-Ni-Sn system reduces the initial stress between the interfaces and also hinders the formation of intermetallic compounds (IMCs). However, these theories fail to explain why whiskers eventually still form when a Ni sublayer is used. Other studies on whisker mitigation methods involving Bi or In were reported very recently [23–25].

In this paper, we examine the idea of replacing Ni with nickel oxide (NiO), as a sublayer for the purposes of mitigating whisker growth. NiO films deposited by vacuum based or other techniques (including the films obtained in this work) are typically slightly non-stoichiometric and semiconducting, and have been used or studied for light emission applications[26], electrochromic devices[27], as a material for resistive random access memory (RRAM) devices[28], and also in contact electrodes for perovskite based solar cells[29]. NiO can be deposited by pulsed laser deposition (PLD)[30], chemical vapor deposition (CVD)[31], sol-gel processes[32], and sputtering[33]. Coating of uniform NiO layer with reasonably good crystallinity[34] *on an industrial scale* can be easily accomplished by the sol-gel method.

## 2. Experimental details

Two types of samples are examined in this study. The first type is referred to as control samples in which Sn is deposited on a Cu coated substrate without a sublayer between Sn and Cu. We call the samples in this category as control samples. In the second type, NiO was deposited onto Cu coated substrates and then Sn was deposited on top of the NiO layer. This type of samples are referred to as NiO sublayer (NSL) samples. In both cases, Cu is used as an under-layer to simulate the Sn coatings in the electronic parts that are either made of Cu or contain Cu conducting tracks (such as in PCB board, etc.).

We used multiple cleaning steps for our substrates. The cleaning procedure is as follows: wash in cleaning solution (Micro-90), then thorough rinse with DI water, followed by ultra-sonication bath in methanol for 20-25 min, and, finally, ultra-sonication bath in ethanol for 20-25 min. In between these steps, the surfaces are rubbed with lint-free wipe and blown dry with nitrogen.

Other relevant parameters and fabrication details are given next. A 200nm (±10nm) thick Cu layer is coated by vacuum evaporation of 99.999% pure Cu (from Kurt J. Lesker) on clean substrates simultaneously. Several Cu coated samples from this group are designated as NSL samples and the others as control samples. NiO layer is then deposited on the NSL samples by reactive radio frequency magnetron sputtering of a 99.99% pure-Ni target in a 20:80 oxygen to argon partial pressure to a thickness of 100nm[35]. Finally, a 250nm (±40nm) Sn layer is deposited on NSL and control samples simultaneously by vacuum evaporation of 99.999% pure Sn (from Kurt J. Lesker) as described elsewhere[36]. The evaporation steps of Cu and Sn were carried out in a Denton Vacuum machine (model DV-502A) at a base vacuum of $(3.5 - 4.8) \times 10^{-6}$ Pa. The NiO deposition is done in a Torr International sputtering machine (model 2G2-TH2).

## 3. Results and discussion

An SEM image of a NiO sublayer (top view) deposited on the Cu layer is shown in Fig 1Figure 1. It indicates a high-quality continuous film growth.

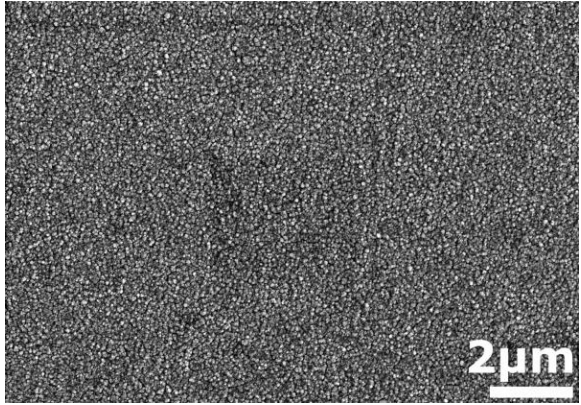

**Figure 1: NiO sublayer on Cu before the deposition of Sn.**

The surface morphology of the samples was examined by scanning electron microscopy (SEM) in a Hitachi S-4800 machine, operated in secondary-electrons mode with acceleration voltage of 5 kV. Compositional analysis was performed using an Oxford Instruments' energy dispersive X-ray spectrograph (EDS), installed in the SEM system. The EDS software (INCA) was calibrated using cobalt EDS standards prior to the EDS point & I.D. mapping. I-V measurements between different layers were performed using Keithley 4200 semiconductor parameter characterization setup. The formation of any intermetallic compounds (IMC) and the thickness of the layers in the control and NSL samples were verified by imaging of cross-section samples obtained by focused ion beam (FIB) milling using a Ga source in a FEI Quanta 3D FEG machine.

The I-V curves, shown in Fig 2, are obtained from two-probe measurements (one probe connected to the top Sn layer surface and the other to the Cu-coated substrate) done on both the control and the NSL samples. These I-V curves are reasonably linear and show good ohmic contacts and behavior in both cases. Since the resulting resistance value of the NSL sample (15.0 m$\Omega$) is very close to the control sample's resistance value (11.3 m$\Omega$), these measurements confirms that the NiO layer did not introduce any significant additional resistance between the Cu and the Sn layers. The NSL sample's cross-section (obtained by FIB milling) was examined over a stretch of more than 20µm using multiple high-resolution SEM images. Fig 3(a) shows a typical and representative NSL sample cross-section. It confirms that a uniform and continuous NiO sublayer between Cu and Sn was obtained. Additionally, a similar cross-sectional analysis is performed on the control sample (see Fig 3(b)) to verify the consistency (of Sn film) and identify any alloy formation between the Cu and Sn. Upon examining Cu-Sn interface also over a stretch of more than 20µm, no evidence of alloys that formed between Cu and Sn over the entire control sample was found, although whiskers have grown right above this region. Also, we did not observe any features between the layers that can be identified as intermetallic compounds (IMCs)[4,14,22]. The formation of IMCs is often considered solely responsible for, or at least contributing to, the whiskers growth although there are contradicting observations[37–39]. EDS analysis was performed to confirm the chemical composition of the deposited films.

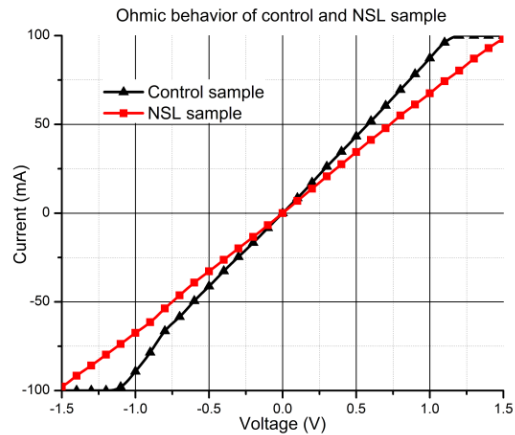

**Figure 2: I-V curves from the control and the NSL samples.**

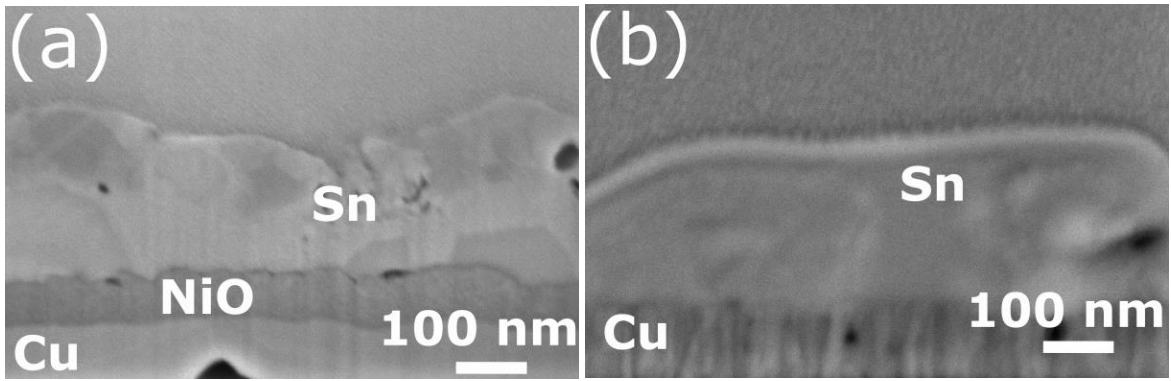

**Figure 3: SEM image showing the cross-section of (a) NSL sample (b) Control sample performed using FIB technique.**

SEM imaging on control and NSL samples was performed immediately after the deposition to verify any initial whisker formation. SEM imaging was then conducted as a function of time, every 15 days or 30 days. Throughout this study, whisker densities were determined by manually counting whiskers from images taken at random over ten equal areas (128 µm x 90 µm). In our study we have observed the formation of whiskers with varying lengths and diameters. Whiskers that are short and thin are less likely to cause any failures in the electronic parts due to their size. Most of the whisker related losses, that have been identified, are caused by the long whiskers[40]. So, we have classified the whiskers based on their lengths. We observed that Class-1 whiskers are more frequent compared to Class-2 and Class-3.

- Class-1: Length up to 2µm.
- Class-2: Length between 2µm and 5µm.
- Class-3: Length greater than 5µm.

Immediately after the samples preparation, our SEM observations showed that there was no detectable whisker growth on either the control or the NSL sample, as can be seen in Fig 4. About 4 weeks later, we found whiskers belonging to Class-1 (Fig 5(a) and Fig 5(c)) on the control sample. No sign of any whisker growth was observed on the NSL sample (Fig 5(b)).

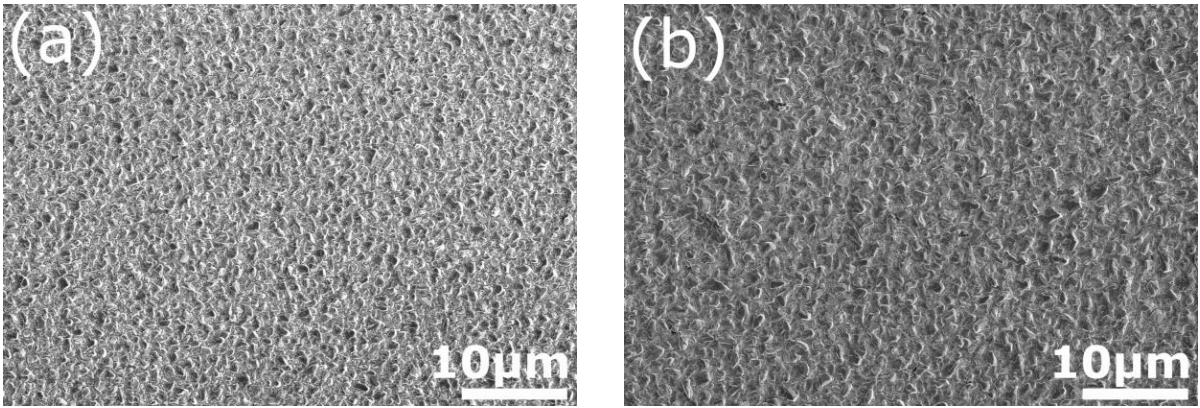

**Figure 4: No whiskers on (a) Control sample and (b) NSL sample.**

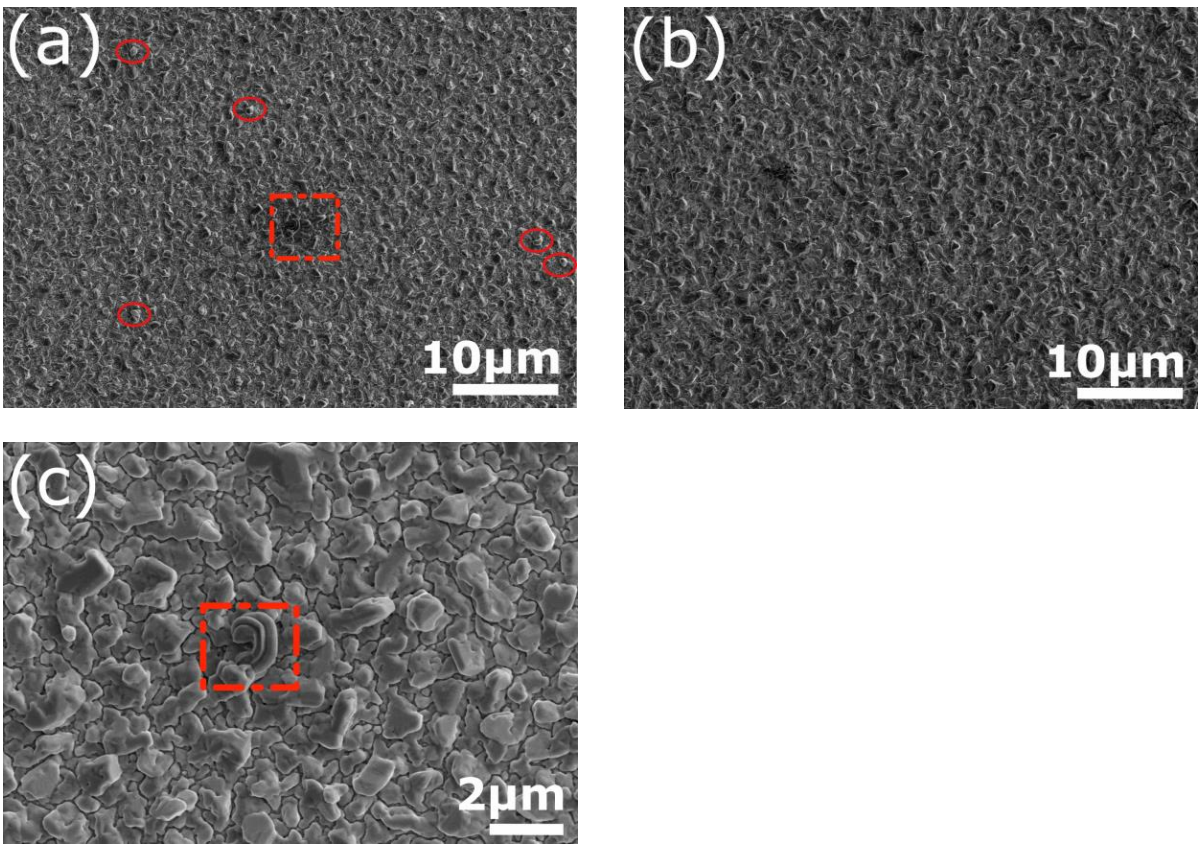

We have continued to follow the samples surface evolution by periodic SEM imaging of the control and NSL samples. The whisker density (whiskers/mm$^2$) calculated over the course of 12 months is summarized in Table 1.

**Table 1. Summary of whisker density (in mm⁻²) in control and NSL sample.**

| Sample type | 2 weeks | | | 6 Months | | | 12 Months | | |
|---|---|---|---|---|---|---|---|---|---|
| | *Class-1* | *Class-2* | *Class-3* | *Class-1* | *Class-2* | *Class-3* | *Class-1* | *Class-2* | *Class-3* |
| **Control sample** | 0 | 0 | 0 | $1848\pm84$ | 0 | 0 | $1673\pm113$ | $1215\pm98$ | $607\pm66$ |
| **NBL sample** | 0 | 0 | 0 | 0 | 0 | 0 | 0 | 0 | 0 |

It can be noted from the above table that, there were only Class-1 whiskers during the first 6 months. As timed passed, ==while new Class-1 whiskers continue to grow, some of the previous Class-1 whiskers appear to have grown further to Class-2 and even Class-3 whiskers.==

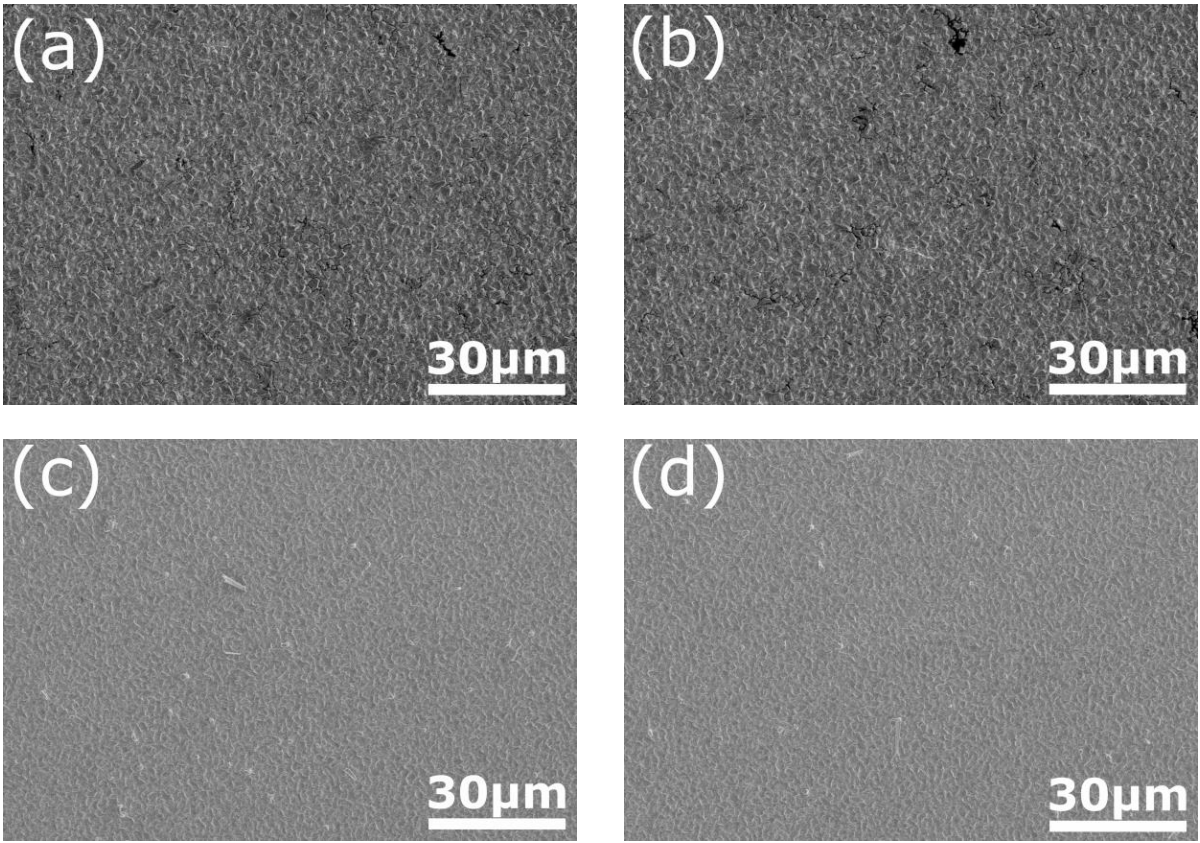

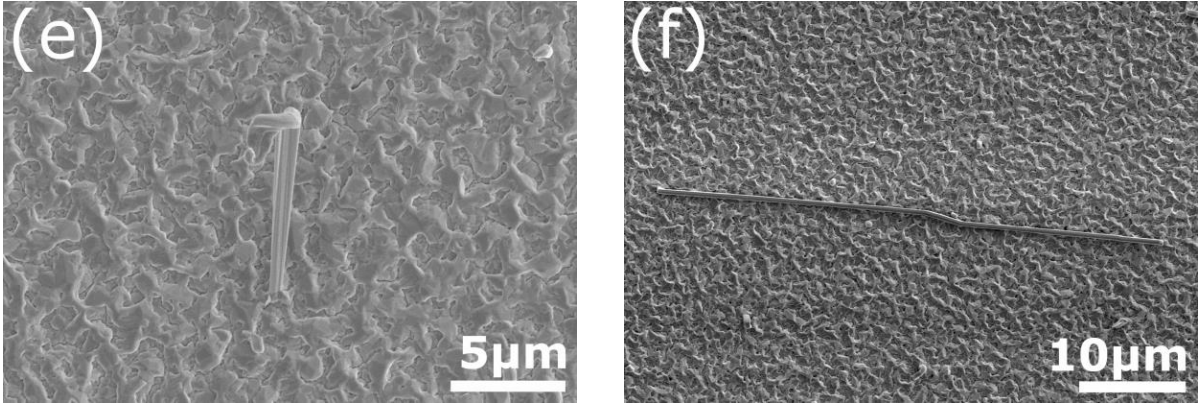

**Figure 6: SEM images (a) and (b) reveal no sign of whiskers on NSL sample, images (c) and (d) details a dense whisker formation on the control sample, this corresponds to the density of ~ 3495±129 whiskers/mm2. (e) lower-magnification image of Class-3 whisker on the control sample after 1 year of incubation. (f) So far the longest whisker on the control sample.**

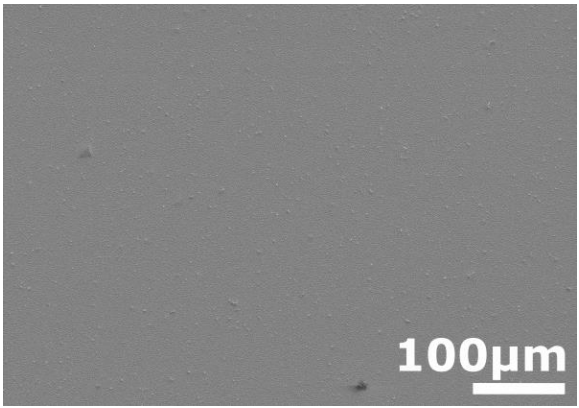

**Figure 7: Very large area scan of control sample representative wide overview showing many whiskers (better seen under higher magnification, such as 800% zoom in the pdf format). (Very HD image)**

The SEM images in Fig 6(c) and Fig 6(d) illustrate whiskers that belong to different classes that have grown on the control sample. The overall whisker density on the control sample has a value close to 3495±129 (whiskers/mm$^2$), this includes all the 3 classes. A close-up view on one such whisker is shown in Fig 6(e), with a length close to 10µm. During our study, the longest whisker that we were able to image on the control sample has a length close to 65µm, shown in Fig 6(f). In contrast to the control sample, no signs of any whisker formation is observed on the NSL sample, shown in Fig 6(a) and Fig 6(b). The large-area high-resolution SEM image in Fig 7 illustrate whiskers with wide range of lengths that have grown on the control sample over an area of ~0.25 mm$^2$.

At this point, we do not appear to have enough information in order to discuss the role of NiO sublayers in, essentially, the elimination of the growth of whiskers. Understanding the mechanism of whisker suppression by using a NiO sublayer, however, would be of significant interest as it may provide clues to what causes the whiskers growth in the first place. It may also facilitate the development of other methods for mitigation or elimination of whiskers growth. The existence of mechanical stress in the Sn film on Cu and a stress reduction in the case of the Sn/NiO/Cu combination is still a possible way to explain the observations in this work. IMC formation at the Sn/Cu interface and a resulting stress build-up is a commonly assumed mechanism of whiskers formation; however our FIB cross-sections work gave no evidence for IMCs formation in either case.

Another scenario, based on the electrostatic theory of whiskers growth [10] is perhaps worth considering. According to that theory, the existence of substantial electric fields can drive the formation of metal whiskers. These electric fields can be externally applied or intrinsic to the materials, the latter attributed to uncompensated electric charges at interfaces, grain boundaries, local variations in chemical compositions, and others [10,41]. For the case under consideration in this work, a simplified electrostatic picture is based on the difference between the work functions of Sn and Cu which is calculated to be ~ 0.53 eV [42,43] [42,43] resulting in a significant built in electric field at the Sn/Cu interface when the two materials do not intermix. It remains unclear how or if that interface-confined field can extend its influence across the Sn film accelerating whisker growth. However the diffusion of Cu into Sn, especially given the grain boundary contributions, will result in a significantly nonuniform (see e. g.[44]) lateral distribution of Cu, whose local concentration will fluctuate from almost zero to almost 100 percent thus leading to the corresponding local variations in work function of the order of the above estimated 0.53 eV. Because the free electrons will redistribute themselves in such a way as to minimize the system free energy, the electric charge distribution becomes laterally nonuniform. Such a nonuniform 'electric patch' structure reaching to the film surface is a major precondition of the electrostatic theory of whisker growth.

We note that the above understanding does not require the existence of IMCs as a driver of whisker growth. Instead it implies that while Cu diffusion is important as it leads to a complementary effect of lateral electric nonuniformities responsible for whisker development. The same understanding can explain whisker suppression in the Sn/NiO/Cu case if we assume that NiO layer blocks Cu diffusion, thus suppressing electric nonuniformities and mitigating whisker growth. As an additional argument, we note that the contact potential between Sn and NiO, which is the approximate work function difference between the two materials, can be as small as ~ 0.4 eV or even less. This estimate is obtained from the electron affinity [45] and band gap [35,46] [35,45] values for NiO assuming almost degenerate p-type NiO[35][35].

A verification of this last set of values would require precise knowledge not only of the values of the work functions but also the effect of any interface layers, contaminations and impurities as well as surface states, which is not available to us at this time. Still, a possible interpretation of these results could be developed in the frame of the electrostatic theory of whiskers [10–12] if some additional experimental results become available.

## 4. Conclusion

In summary, (1) we observed that NiO sublayer has a very strong impact on suppressing the whisker formation, (2) the almost negligible resistance that the NiO sublayer adds to the Sn layer, combined with the possibility for a low-cost application methods of NiO, such as sol-gel or other solution based methods, suggest significant potential in applications.

### Acknowledgments

This work has benefitted from technical help with the deposition of films, provided by the Wright Center for Photovoltaics Innovation and Commercialization (PVIC) at the University of Toledo. We are also thankful to Dr. Victor Karpov from the Department of Physics and Astronomy, University of Toledo, for advice and insightful discussions on the topic.